\documentclass[aps,english,prb,twocolumn,showpacs,superscriptaddress]{revtex4}
\usepackage{graphicx}
\usepackage{amssymb}
\usepackage{babel}

\begin{document}

\title{Current-voltage (I-V) characteristics of armchair graphene nanoribbons under uniaxial strain}

\author{M. Topsakal}
\affiliation{UNAM-Institute of Materials Science and Nanotechnology, Bilkent University, Ankara 06800, Turkey}
\author{V. M. K. Bagci}
\affiliation{Research Center for Applied Sciences, Academia Sinica, Taipei 115, Taiwan}
\author{S. Ciraci}\email{ciraci@fen.bilkent.edu.tr}
\affiliation{UNAM-Institute of Materials Science and
Nanotechnology, Bilkent University, Ankara 06800, Turkey}
\affiliation{Department of Physics, Bilkent University, Ankara
06800, Turkey}

\date{\today}

\begin{abstract}
The current-voltage (I-V) characteristics of armchair graphene
nanoribbons under a local uniaxial tension are investigated by
using first principles quantum transport calculations. It is shown
that for a given value of  bias-voltage, the resulting current
depends strongly on the applied tension. The observed trends are
explained by means of changes in the band gaps of the nanoribbons
due to the applied uniaxial tension. In the course of plastic deformation, the irreversible structural changes and derivation of carbon monatomic chains from graphene
pieces can be monitored by two-probe transport measurements.
\end{abstract}

\maketitle

\section{Introduction}

Graphene, as a (2D) monolayer honeycomb structure of carbon, has attracted a great deal of interest since its successful preparation in 2004.\cite{gra2004} Due to its unique mechanical, structural and electronic properties, graphene have been realised as an important material for numerous theoretical investigations and promising applications. Among these are charge carriers behaving as massless Dirac fermions,\cite{dirac} Klein tunneling,\cite{klein1,klein2} ballistic transport at room temperature,\cite{ballistic1,ballistic2}  and anomalous quantum Hall effects.\cite{qhe} From experimental points of view, field-effect transistors\cite{transistor1,transistor2}, micromechanical resonators\cite{resonator}, gas sensors\cite{gas_sensor} of graphene have already been proposed. Most of these are directly related with its transport properties. 

Earlier transport studies predict that spin-valve devices based on graphene nanoribbons can exhibit magnetoresistance values that are thousands of times higher than previously reported experimental values.\cite{magnetorezistance} Unusual effects of dopings on the transport properties of graphene nanoribbons were also reported.\cite{doping1,doping2,hasan} Nevertheless, the transport properties of graphene nanoribbons under uniaxial tension have not been fully explored even from the theoretical points of view. While the effect of strain on the electronic properties of graphene is becoming an active field of study,\cite{neto} the transport properties and I-V characteristics of nanoribbons under local or uniform strain is of crucial interest for development of future device applications.

In this study, based on state-of-the-art first-principles quantum
transport calculations, we investigate the effects of uniaxial
strain on the current-voltage (I-V) characteristics of graphene
nanoribbons. We showed that elastic strain can alter the electron
transport properties dramatically. In some cases, under a 10\%
strain, the current can change as much as 400-500 \%. However,
the variation of current with strain is sample specific. Even more
remarkable is that the chain formation of carbon atoms from the
graphene nanoribbons\cite{iijima,mehmet_kopma} undergoing a
plastic deformation can be monitored through I-V characteristics
showing negative differential resistance.

\section{MODEL AND METHODOLOGY}
Geometry relaxations and electronic structures are calculated by
using SIESTA  package,\cite{siesta} which uses
numerical atomic orbitals as basis sets and
Troullier-Martin type\cite{TM} norm-conserving
pseudopotentials. The exchange-correlation functional of the
generalized gradient approximation is represented by the
Perdew-Burke-Ernzerhof approximation.\cite{pbe} A 300 Ryd mesh
cut-off is chosen and the self-consistent calculations are
performed with a mixing rate of 0.1. The convergence criterion for
the density matrix is taken as 10$^{-4}$. Brillouin
zone (BZ) sampling of the calculations have been determined after
extensive convergence analysis. The conjugate gradient
method is used to relax all the atoms until the maximum absolute
force was less than 0.05 (eV/\AA). Interactions between adjacent
graphene layers is hindered by a large spacing of $\sim$10 \AA.

The electronic transport properties are studied by the
non-equilibrium Green's function (NEGF) techniques, within the
Keldysh formalism, based on density functional theory (DFT) as
implemented in the TranSIESTA (Ref. \onlinecite{transiesta}) module within the SIESTA (Ref. \onlinecite{siesta}) package. A  single $\zeta$-plus-polarization basis set is used. Test
calculations with larger basis set and mesh cut-off were also
performed, which give almost identical results. The current
through the contact region was calculated using Landauer-Buttiker
formula,\cite{buttiker}
\begin{equation}
I(V_b)=G_0\int_{\mu_R}^{\mu_L}T(E,V_{b})dE,
\end{equation}
where \textit{G$_0=2(e{^2}/h$)} is the unit of quantum conductance  and
$T(E,V_{b})$ is the transmission probability of electrons incident
at an energy \textit{E} through the device under the potential
bias $V_b$.  The  electrochemical potential difference between the
left and right electrodes  is $eV_{b} = \mu_L-\mu_R $.

\section{I-V characteristics under elastic strain}
The band gaps of armchair graphene nanoribbons (AGNRs), which we
consider in this study, depend on their widths,\cite{cohen_prl}
which are conventionally specified according to the number of
dimer lines, $N$ in their primitive unit cell. AGNR($N$)'s are
grouped into three families, namely $N=3m-1$ family having smallest
band gaps, the $N=3m$ family having medium band gaps, and
$N=3m+1$ family having largest gaps, where $m$ is an integer. Band
gaps of each family decrease with increasing $m$ and eventually
goes to zero as $m \rightarrow \infty$. AGNRs are nonmagnetic
direct band gap semiconductors, which can, however, be modified by
vacancies\cite{mehmet_delik} and impurities.\cite{haldun_tm}

\begin{figure}
\includegraphics[width=7cm]{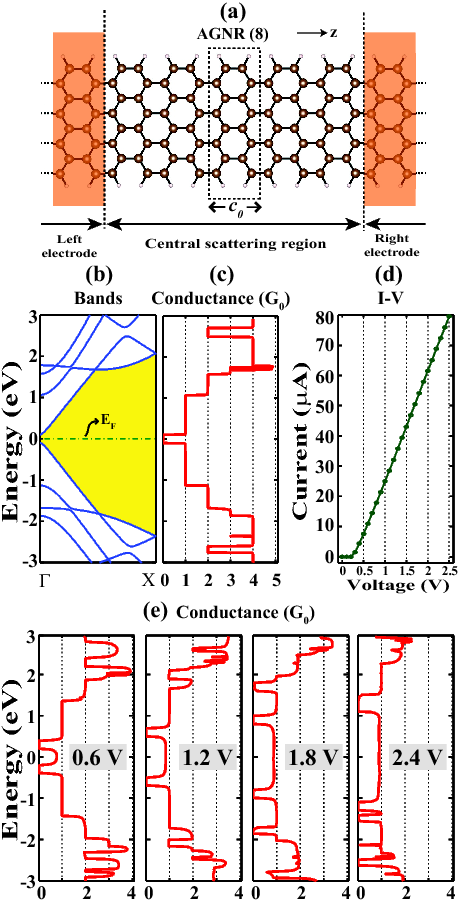}
\caption{(Color Online) (a) Schematic view of two-probe armchair
graphene nanoribbon system AGNR(8), having $N$=8 dimers along the
$z$-axis. Central scattering region, left and right electrodes are
indicated. Carbons atoms are represented by large (brown) and H
atoms by small (light) balls. Primitive unit cell of electrodes
and scattering region are the same and represented by dashed lines.
The lattice constant of the primitive unit cell is $c_0$, and that of 
the central scattering region is $5c_0$.
(b) Band structure of AGNR(8) in its primitive unit cell. (c)
Transmission spectrum of the system shown in (a) under zero-bias
voltage. (d) I-V plot of AGNR(8) for a bias voltage
from 0 to 2.5 V. (e) The transmission spectrums for 4 different bias 
voltage calculated for the system shown in (a).} \label{fig:1}
\end{figure}

The NEGF technique used to study the electronic transport employs
a two-probe system; semi-infinite left- and right-electrode
regions are in contact with a confined central scattering region. A
two-probe system, specific to AGNR with $N=8$, but representative
of any $N$, is shown in Fig.~\ref{fig:1} (a). Both electrodes and
the central region are made from AGNR($8$). Periodic boundary
conditions were imposed on the plane perpendicular to the axis of
the nanoribbon. The carbon atoms at the edges are saturated with H
atoms. The central region contains $5$ primitive unit cells, with
a total length of 21.76 \AA{}  ($=5c_{0}$). The length of the central
region is sufficient enough to avoid an abrupt change in
electronic structure while progressing from the electrode region
to the strained region of interest.

We first consider the electronic transport properties of the
unstrained two-probe system presented in Fig.~\ref{fig:1} (a). To
provide for an intuitive understanding of the transport phenomena,
the band structure of the electrodes or the scattering region in
their primitive unit cell are shown in Fig.~\ref{fig:1} (b). The
lowest conduction and highest valance bands originate from
$\pi^*$- and $\pi$-states, respectively. Unlike the perfect 2D
graphene, where $\pi$- and $\pi^*$-bands cross at the $K$-corners
of BZ, AGNR(8) is a direct band gap material having
0.20 eV band gap value. The calculated zero-bias transmission
spectrum is given in Fig.~\ref{fig:1} (c), which apparently mimics
the band structure of AGNR(8). There is a region of
zero-transmission with a width of $0.20$ eV and located around the
Fermi level, coinciding with the band gap of AGNR(8). Likewise,
the step-like behavior of the spectrum is related with the
available conductance channels due to bands.  The current as a
function of the applied bias voltage $V_b$ is presented in
Fig.~\ref{fig:1} (d). For this type of calculations, we increased
$V_b$ in steps of 0.1 V and used the converged density matrix of
the previous state as an initial guess for the next step. Applying
a bias voltage shifts the Fermi level of the left-electrode with
respect to the Fermi level of the right-electrode. The current
starts flowing once the top of the valence band of the
left-electrode matches in energy with the bottom of the conduction
band of the right-electrode, as expected from the \textit{T(E, $V_{b}=0$)}
for low-bias values. The \textit{T(E,$V_{b}$)} does not alter much with the
bias, since it is a uniform system and no significant permanent
charge migrations should occur. This is evident from the linear
response of the current to the bias voltage for values of
$V_{b}>0.2$ eV. As the calculation of the current is very time
consuming, the bias range is limited from 0 to 2.5 V. In Fig.~\ref{fig:1} (e)
the transmission spectrums for bias voltages of 0.6 V, 1.2 V, 1.8 V and 2.4 V 
are also presented. As readily seen, the transmission T(E,$V_{b}$) contributing to 
the current always keep near 1 \textit{G$_0$}, and the higher values in transmission values 
move further away from the Fermi energy. This is due to the fact
that as the bands of the leads move up or down in energy scale with 
the varying bias, only a single conduction channel is open, or in other words, 
only one band crosses the energy of interest, at either one or both of 
the leads. This holds true for the bias voltages we consider in this study 
and as a result, we see a linear current response to voltage 
for zero-strain.

\begin{figure}
\includegraphics[width=7cm]{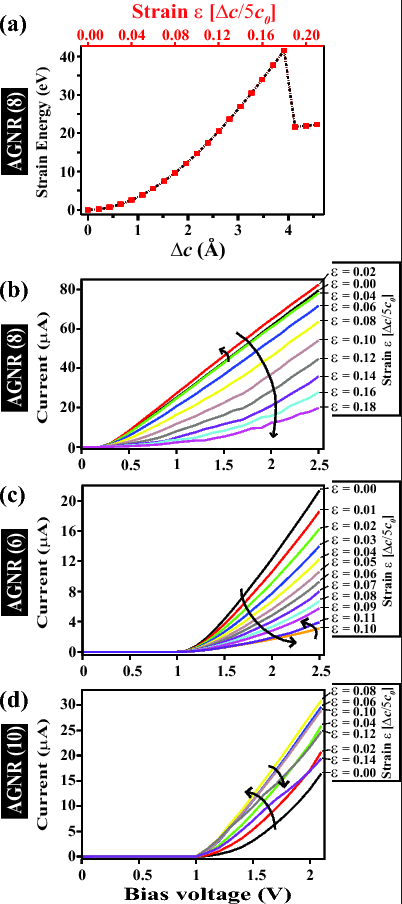}
\caption{(Color Online) (a) The strain energy $E_{S}$, as a
function of elongation $\Delta c$, or strain $\epsilon=\Delta
c/c_{0}$ for AGNR(8) system. (b) I-V
characteristics of AGNR(8) for a bias voltage from 0 to 2.5 V for
different values of strain. The increase and decrease trends are
shown by arrows. (c) and (d) same as (b) for different systems,
AGNR(6) and AGNR (10), respectively.} \label{fig:2}
\end{figure}

Earlier, we have investigated the elastic and plastic deformation
of graphene and its nanoribbons under uniaxial
tension.\cite{mehmet_kopma} Mechanical properties were revealed
from the strain energy, $E_{S}=E_{T}(\epsilon)-E_{T}(\epsilon=0)$;
namely, the total energy at a given uniaxial strain $\epsilon$
minus the total energy at zero-strain. Here, the uniaxial strain
is $\epsilon=\Delta c/c_{0}$, where $c_{0}$ and $c=c_{0}+\Delta c$
are equilibrium and stretched lattice constants of the nanoribbon, respectively.
The tension force, $F_{T}=-\partial E_{S}(\epsilon)/\partial c$
and the force constant $\kappa=\partial^{2}E_{S}/\partial c^{2}$
are obtained from the strain energy. Calculated elastic constants
were in good agreement with available experimental data obtained
from graphene.\cite{gr_exp} Here we consider the I-V
characteristics of AGNR(8) under a uniform uniaxial tension of the
central scattering region for $0\leq \epsilon \leq 0.18$. The
strain is introduced as follows: The electrode atoms are fixed in
their equilibrium positions while the length of the central region
is increased uniformly by $\Delta c$. Subsequently, the structure
of the central region is fully optimized in a larger
supercell containing also unstrained electrode regions. In this
respect, our study reveals the effect of a local strain in a long
unstrained nanoribbon. The total energy of the system is
recalculated. The strain energy $E_{S}$, is obtained according to
above definition. $E_{S}$ versus the elongation $\Delta c$, as
well as $\epsilon = \Delta c/5c_{0}$ plot for AGNR(8) system is given in
Fig.~\ref{fig:2} (a). The segment of AGNR(8) in the central
scattering region undergoes an elastic deformation up to strain
values $\epsilon\simeq 0.18$, where the honeycomb-like structure
is maintained, and the system returns to its original
configuration if the tension is released. However, for higher
values of strain, the system deforms plastically, where
irreversible structural changes occur and the strain energy
suddenly drops. Further information for this type of elastic and
plastic deformation can be found in Ref.\onlinecite{mehmet_kopma}.

\begin{figure}
\includegraphics[width=7cm]{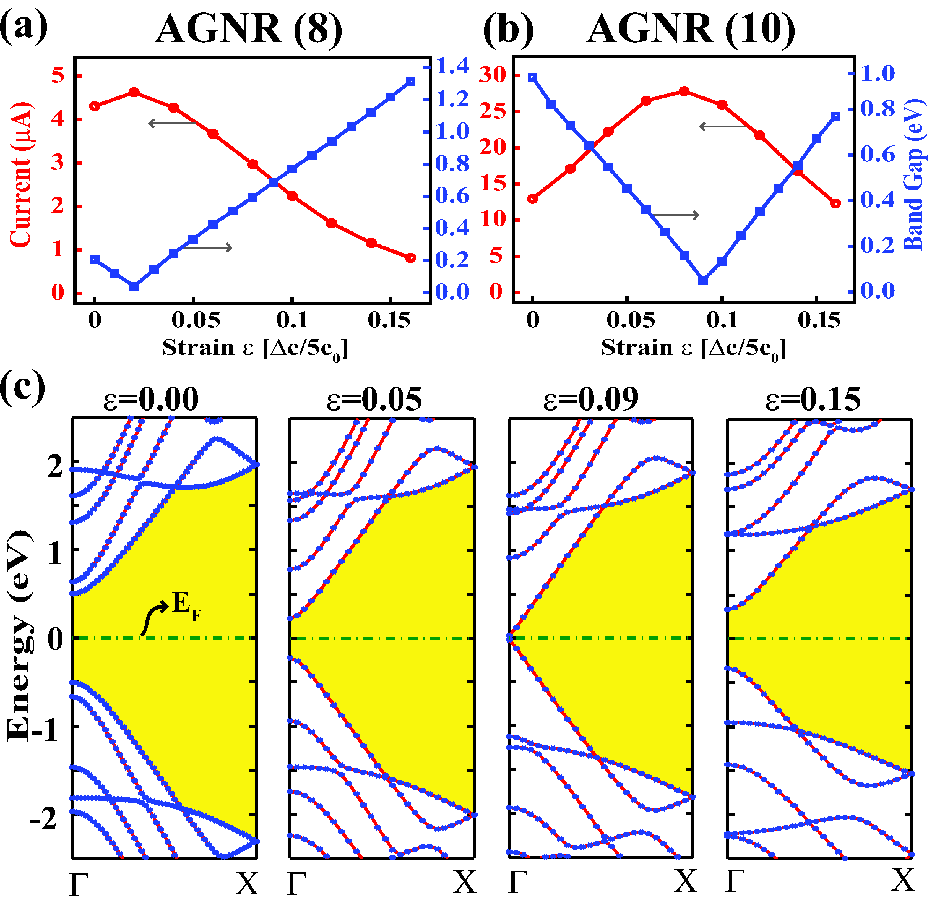}
\caption{(Color Online) The values of band gaps and currents of
(a) AGNR(8) and (b) AGNR(10) systems calculated as a function of
strain, $\epsilon$. The currents are calculated for 2.0 V
bias-voltage. (c) The band structures of AGNR(10) under different
strains. The band gaps are shaded and Fermi energy is set to
zero.} \label{fig:3}
\end{figure}

In Fig.~\ref{fig:2} (b)-(d), we present the I-V plots of stretched
nanoribbons. The electrode regions are identical to the no-strain
case, but the central region under strain causes the changes. Once
again, due to the band gap of the electrodes of 0.2 eV in AGNR(8), no current
is observed up to a bias voltage of 0.2 V. The current response to
bias voltage is linear for low strain, but becomes increasingly
non-linear for higher strain. It is important to notice that
higher strain in the central region induces stronger
non-uniformity on its geometry and thus on its electronic
structure as compared to the electrodes. Equilibrium charge
transfer may occur and alter the systems response to the
non-equilibrium perturbation. This will result in a varying
\textit{T(E,$V_{b}$)} for different values of $V_{b}$. It is informative
to compare the current values for systems under different strain
at a given bias voltage. For example, at $V_{b}$=2 V, the current
is around 62 $\mu A$ at $\epsilon=0$. It increases to 65 $\mu A$
at $\epsilon=0.02$, but steadily decreases for higher strain,
having values of 28 $\mu A$ at $\epsilon=0.12$ and 12 $\mu A$ at
$\epsilon=0.18$. One also notes that the I-V curve in
Fig.~\ref{fig:1} (d), which is almost linear for $\epsilon=0$,
starts to lose its linearity for higher values of strain as seen
from Fig.~\ref{fig:2} (b).

Other ribbons such as AGNR(6) and AGNR(10) whose I-V
characteristics are given in  Fig.~\ref{fig:2} (c) and (d).
AGNR(6) belongs to the $N=3m$ family and it has a larger band gap
($\simeq$1.04 eV) as compared to AGNR(8). As a result, we do not
observe a current until 1 eV as seen from Fig.~\ref{fig:2} (c).
The current steadily decreases until $\epsilon=0.10$, then starts
to increase as seen from Fig.~\ref{fig:2} (c). AGNR(10) is another
system which has a band gap value around 1.00 eV and its I-V characteristics are given in Fig.~\ref{fig:2}
(d). In contrast to AGNR(6), the current first increases until
$\epsilon=0.08$ and then starts to decrease for higher strain
values. All these results in Fig.~\ref{fig:2} (b),(c), and (d)
show that the current passing through nanoribbons is very
sensitive to the strain values and the behaviors of I-V curves are
sample specific.

\begin{figure*}
\includegraphics[width=17cm]{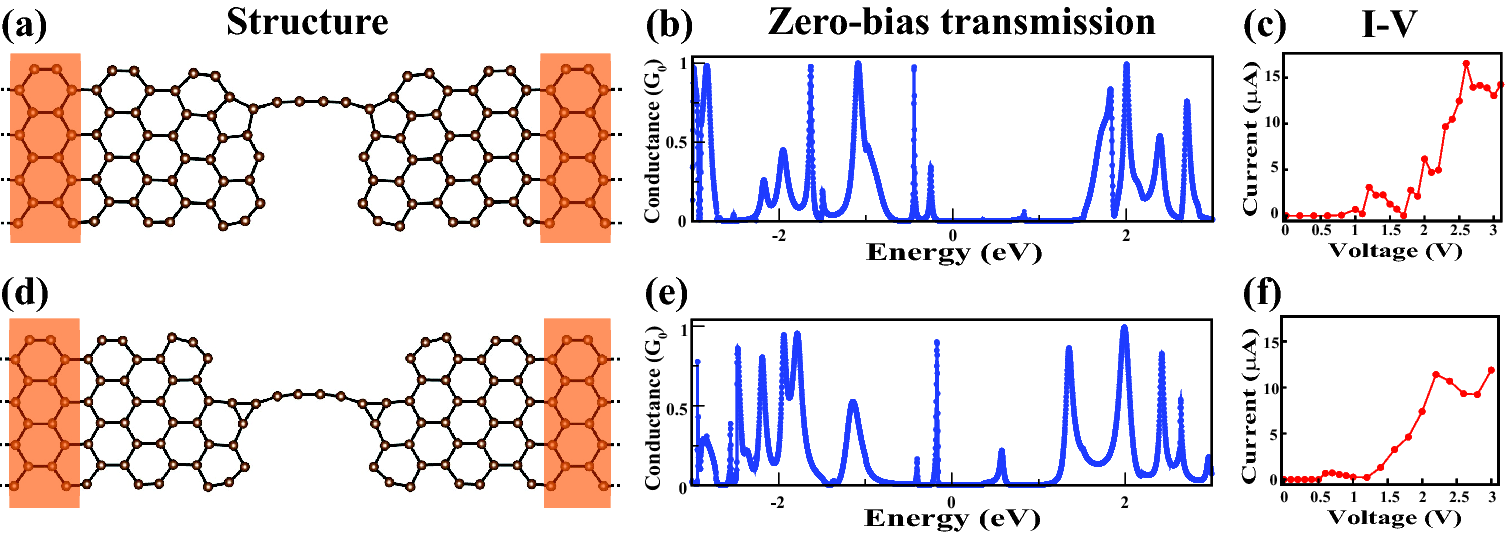}
\caption{(Color Online) (a)  Schematic view of AGNR(8) system
which is deformed plastically due to the high values of strain
($\epsilon > 0.20$). Its zero-bias transmission (b), and I-V characteristics
(c), are given on the right. (d)-(e)-(f) are the same as
(a)-(b)-(c) for a longer chain between graphene pieces.}
\label{fig:4}
\end{figure*}

The increase and decrease of the currents given in
Fig.~\ref{fig:2} due to the changes in the strain is directly
related with the electronic structure of the central scattering
region, which is modified as a result of changes in atomic
structure under tension.\cite{mehmet_kopma,hakem} In Fig.~\ref{fig:3} (a) we show the
variation of current and band gap of AGNR(8) with applied uniaxial
strain. Here current values are extracted from Fig.~\ref{fig:2}
(b) for 2 V bias voltage. As seen from the plots, there is an
inverse relationship between the current and band gap values. Any
increase in the band gap decreases the current and vice versa. The
same analysis performed for AGNR(10) in Fig.~\ref{fig:3} (b) also
confirms this relationship. The changes in the band structures of
AGNR(10) can also be followed from Fig.~\ref{fig:3} (c), where the
lowest conduction and highest valence bands approach to each other
until $\epsilon\simeq0.09$ and then move away for higher values of
strain. The band gap variations occur due to different nature of bands
around the conduction and valence band edges exhibiting different
shifts with strain. In particular, note that $\pi^*$- and
$\pi$-bands of AGNR(10) cross linearly at $\Gamma$-point by
closing the band gap. This is the realization of massless Dirac
Fermion behavior in a nanoribbon, which is semiconductor under zero-strain.\cite{mehmet_kopma} In a
simple model, an electron is ejected from the left electrode at an
energy value lower than the shifted valence band maximum, for
available ranges within the bias voltage. It is incident upon the
central region, with lower chemical potential, and tunnels through
to the right electrode, still lower in energy. The smaller the
band gap value for the central region, the larger the number of
possible states that participate in the tunneling, thus the
larger is the value of the current.

\section{Transport properties of AGNR(8) under plastic deformation}
While the elastic deformation imposes changes in the band gap and
current values, the onset of plastic deformation results in
dramatic changes in the structure. After yielding, the
modification of honeycomb structure is somehow stochastic and
sample specific. It depends on the conditions, such as the defects
in the sample, the temperature effects and the rate of stretching.
However, it has been shown theoretically\cite{mehmet_kopma} that
under certain circumstances a long carbon atomic
chain\cite{tongay} (identified as cumulene having double bonds and
polyyne with alternating triple and single bonds) can form in the
course of plastic deformation of graphene, unless the edges of
AGNR is not terminated with hydrogen. Upon further stretching,
each carbon atom of graphene implemented to chain results in a
stepwise elongation of the chain between two graphene pieces.
Monatomic carbon chain, which was derived experimentally from
graphene,\cite{iijima} can be a potential nanostructure for
various future applications. The important issue we address here
is how these sequential structural changes reflects the transport
properties.

In Fig.~\ref{fig:4} (a) we present the atomic structure of a
two-probe graphene nanoribbon system which is formed after the
plastic deformation of AGNR(8) nanoribbon. A short chain
containing 4 carbon atoms between the graphene flakes is formed in
the scattering region and its zero-bias transmission spectrum is
presented in Fig.~\ref{fig:4} (b). This spectrum is composed of
peaks rather than step-like levels as in Fig.~\ref{fig:1} (c). The
calculated I-V plot in Fig.~\ref{fig:4} (c) also contains some peaks,
which may lead to negative differential resistance.\cite{cohen_differential} Similar situation also
exists for a longer chain in Fig.~\ref{fig:4} (d), which occurred
at a more advanced stage of plastic deformation whereby the
nanoribbon in the central region is more stretched than in
Fig.~\ref{fig:4} (a). At the end, two more carbon atoms are 
included to the chain. The differences between zero-bias
transmission curves in Fig.~\ref{fig:4}(b) and (e) occur because
of the energy level diagram and their positions relative to Fermi
levels are different. Also, the I-V curve corresponding to two
carbon chains of different lengths occurring in subsequent stages
of stretching are rather different. We note that the conductance of the
longer carbon chain in Fig.~\ref{fig:4}(d) and the corresponding current values (I) of a given 
bias voltage (V) can be higher than the shorter chain in Fig.~\ref{fig:4}(a). 
This paradoxical situation is related with the fact that some energies of the channels can be closer to the
Fermi level as the length of the chain increases.\cite{lang} Further stretching of the system
shown in Fig.~\ref{fig:4}(d) can produce longer carbon chain structures.
The length of these chains can be as long as 10 carbon chains. As found for the structures  in Fig.~\ref{fig:4}(a) and Fig.~\ref{fig:4}(d), the I-V characteristics of the longer carbon chains will be different and will allow one to monitor the structural changes. Finally the plastic deformation
terminates upon breaking of the chain.

\section{Conclusion}
We have shown that the transport properties of the
segment of an armchair graphene nanoribbon in a two-probe system
can be modified with uniaxial strain. The current under a fixed
bias can change several times with applied uniaxial strain.
However, these changes are sample specific and related with strain
induced changes in the electronic structure near the band gap.
Irreversible structural changes and the formation of monatomic
carbon chain between graphene pieces in the advanced stages of
plastic deformation can be monitored through two-probe transport
experiments. We believe that our findings are of crucial
importance for recent active studies aiming to reveal the effects
of strain on the electronic properties of graphene. Also our
results suggest that these systems can be used as nanoscale strain
gauge devices.

\section{Acknowledgement}
Part of the computations have been provided by UYBHM at Istanbul
Technical University through a Grant No. 2-024-2007. S.C.
acknowledges financial support from The Academy of Science of
Turkey (TUBA).

\end{document}